\documentclass[doublecol,sd,amsmath,amssymb,amstext]{epl2}
\usepackage{color}
\usepackage{graphicx}
\usepackage{graphicx,epstopdf}
\usepackage{dcolumn}
\usepackage{bm}
\usepackage{mathrsfs}
\usepackage{amsfonts}
\usepackage{amsmath}

\usepackage{lineno}

\begin{document}


\title{Dynamic interaction induced explosive death}

	\author{ Shiva Dixit$^1$, Sayantan Nag Chowdhury$^2$,  Dibakar Ghosh$^2$, and Manish Dev Shrimali$^1$ \footnote{Email: shrimali@curaj.ac.in}}
	\shortauthor{S. Dixit, S. Nag Chowdhury, D. Ghosh and M. D. Shrimali}
	
\institute{$^1$Department of Physics, Central University of Rajasthan, NH-8, Bandar Sindri, Ajmer 305 817, India\\
	$^2$Physics and Applied Mathematics Unit, Indian Statistical Institute, 203 B. T. Road, Kolkata 700108, India

}

\date{\today}

\abstract{Most previous studies on coupled dynamical systems assume that all interactions between oscillators take place uniformly in time, but in reality, this does not necessarily reflect the usual scenario. The heterogeneity in the timings of such interactions strongly influences the dynamical processes. Here, we introduce a time-evolving state-space dependent coupling among an ensemble of identical coupled oscillators, where individual units are interacting only when the mean state of the system lies within a certain proximity of the phase space. They interact globally with mean-field diffusive coupling in a certain vicinity and behave like uncoupled oscillators with self-feedback in the remaining complementary subspace. Interestingly due to this occasional interaction, we find that the system shows an abrupt explosive transition from oscillatory to death state.  Further, in the explosive death transitions, the oscillatory state and the death state coexist over a range of coupling strengths near the transition point. We explore our claim using  Van der pol, FitzHugh–Nagumo and Lorenz oscillators with dynamic mean field interaction. The dynamic interaction mechanism can explain sudden suppression of oscillations and concurrence of oscillatory and steady state in biological as well as technical systems.}

 {Nonlinear dynamics and nonlinear dynamical systems}
  {Interdisciplinary applications of physics}

  {Structures and organization in complex systems}



\maketitle

\section{Introduction} \label{introduction}

\par Explosive transition \cite{boccaletti2016explosive} in ensembles of coupled dynamical systems grabs the attention of many physicists due to its relevance in various practical applications. The transition from incoherence to coherence \cite{boccaletti2002synchronization,chowdhury2019convergence} and the emergence of a giant connected component in the network \cite{bollobas2006percolation} in most of the cases are continuous and reversible. However, abrupt emergence of a collective state due to non-trivial interactions among coupled dynamical systems has also been widely reported \cite{ji2013cluster,kuramoto2003chemical,gomez2011explosive,leyva2012explosive,pazo2005thermodynamic}. The vast majority of these studies are concerned with the discontinuous synchronization transition (also known as explosive synchronization) \cite{zhang2013explosive,khanra2018explosive,jalan2019explosive}, where the synchronization order parameter exhibits an irreversible transition with respect to the varying control parameter. Recently, explosive transition from incoherent dynamics to frequency-locked state is observed in heterogeneous Kuramoto models using attractive and repulsive interactions \cite{frolov2020chimera,majhi2020perspective}.

\par Besides, a new phenomenon of explosive oscillation quenching, termed as explosive death \cite{bi2014explosive,verma2017explosivescirep,verma2018first,verma2019explosive,verma2019explosivepre}, is recently found to provide a rich playground that can be explored successfully with the interdisciplinary approaches of complex systems. Despite its youth, this discontinuous and irreversibile transition during suppression of oscillation is enjoying widespread recognition. During this type of first order like transition, a mixed regime consisting of the oscillatory state and the death state is found to coexist near the transition point. 
 The coexistence of these two states is ubiquitous in many physical systems \cite{herrero2000experimental,wei2007amplitude}, chemical systems \cite{crowley1989experimental,bar1985stable}, biological systems \cite{beuter2003nonlinear} and in several numerical studies \cite{liu2012inhomogeneous,liu2005partial} consisting of limit cycles and chaotic oscillators. Nevertheless, all these percieved studies on explosive death \cite{bi2014explosive,verma2017explosivescirep,verma2018first,verma2019explosive,verma2019explosivepre} are done using static network formalism, where the interactions among those oscillators are assumed to be invariant for the entire course of time.

\par Here, a dynamic coupling configuration \cite{2021arXiv210104005D} is employed in this letter to inspect the explosive death phenomenon. Units in nature remain rarely isolated and the interaction among those units are continuously updating depending on various proximities. Information diffusion over communication networks, data-packet transmission on the web, disease contagion on the social network of patients are perhaps few potential examples, which attests the fundamental necessity of the temporal network approach \cite{holme2012temporal,nag2020cooperation}. Such time-varying interactions among coupled oscillators give rise to fascinating collective phenomenon \cite{rakshit2020intralayer,aprasad_pramana,rakshit2018synchronization, yadav2017dynamics, sudhashu2019,sd1,rakshit2018emergence,schrodertransient, threshold_2019}. Earlier, Majhi et al.\ \cite{majhi2017amplitude} reported the emergence of death state in a temporal network of mobile oscillators. But, the focus of our letter is completely different from that Ref.\ \cite{majhi2017amplitude}. In this letter, we present the first comprehensive analysis of the effects of dynamic coupling configuration on the explosive death transition in an ensemble of identical coupled self-excited oscillators \cite{rajagopal2019camo,nag2020hidden,munoz2018new}. This kind of discontinuous transition from the oscillatory state to the death state using dynamic interaction is reported here for the first time to the best of our knowledge. Moreover in case of mobile agents' network \cite{majhi2017amplitude,chowdhury2019extreme,9170822}, although the mobility of mobile agents affects the collective dynamics of the system, but the states of those oscillators situated on the top of those agents usually do not influence the agents’ mobility. To get rid of this unidirectional affair, a dynamic coupling scheme \cite{2021arXiv210104005D} is implemented here, where the mean state of all oscillators decides whether the interaction among those oscillators will appear or not.

\par The rest of the paper is organized as follows: In the next section, we describe our model and details of the proposed dynamic coupling mechanism. After that, we systematically investigate the transitions between the oscillatory state and the death state in a system of coupled oscillators. To characterize the first-order like transition, we use the normalized average amplitude \cite{sharma2012amplitude} as a measure. To validate our claims, three different dynamical systems (i) Van der Pol oscillator \cite{kanamaru2007van}, (ii) FitzHugh-Nagumo model \cite{fitzhugh1961impulses}, and (iii) Lorenz oscillator \cite{lorenz1963deterministic} are considered as paradigmatic models. We also inspect numerically that how the explosive death transition depends upon control parameters of the systems. Using linear stability analysis, the backward transition point for this explosive transition has been calculated. This backward critical coupling strength does not depend on the size of the system and agrees completely with the numerics. Finally, we summarize our results and conclude.

\section{Mathematical framework}\label{DynInt}

\par Here, $N (\ge 2)$ identical nonlinear oscillators are considered and we couple them via dynamic coupling. Each $j$-th oscillator $(j=1,2,...,N)$ is evolved by the set of differential equations \cite{2021arXiv210104005D},
\begin{equation}
\begin{split}
\dot{\mathbf{X}}_{j} &={F}(\mathbf{X}_{j})+ \varepsilon \beta ({H} \overline{\mathbf{X}} - \mathbf{X}_{j}),
\end{split}
\label{eq1}
\end{equation}
where $\dot{\mathbf{X}}_{j}={F}(\mathbf{X}_{j})$ reflects an isolated dynamics of $j$-th oscillator and $\dot{\mathbf{X}}$ denotes differentiation of $\mathbf{X}$ with respect to time $t$. ${F}(\mathbf{X}_{j}): \mathbb{R}^d \to \mathbb{R}^d$ is the vector field corresponding to the $d$-dimensional vector ${\mathbf{X}}_{j}$ of the dynamical variables. $\varepsilon \ge 0$ is the coupling strength and {$\beta=$diag$(\beta_1,\beta_2,\cdots,\beta_d)$ is a $d \times d$ diagonal matrix, where depending on $\beta_j$ $(j=1,2,\cdots,d)$, the components of $X_j$ take part in the dynamic coupling. $\overline{\mathbf{X}}= \frac{1}{N} \sum_{j=1}^{N} \mathbf{X}_j$ denotes the arithmetic mean of the state variables. The dynamic nature of the coupling mechanism is contemplated through the introduction of the step function $H$. This function $H$ is chosen here as a function of the mean field term $\overline{\mathbf{X}}$. Whenever the mean field term $\overline{\mathbf{X}}$ lies within a pre-specified subset $M \subseteq \mathbb{R}^d$ of the state space $\mathbb{R}^d$, the mean field interaction is activated by setting $H=1$. On the other hand, if $\overline{\mathbf{X}}$ lies in the complementary subset $\mathbb{R}^d \setminus M$, then the mean field interaction is turned off by setting $H=0$. Thus, the range of $H$ consists only two distinct values $1$ and $0$ solely depending on the value of $\overline{\mathbf{X}}$. {In absence of mean field interaction, the oscillators become uncoupled with negative self-feedback only. Interestingly, the oscillators are either completely independent of each other (when $H=0$) or they are globally coupled with each other (when $H=1$). Hence, representing each oscillator as a node and their interactions as edges \cite{threshold_2019}, we have a time-varying network where the degree of each node at each time step is either $0$ (when $H=0$) or $N-1$ (when $H=1$).

\par The subset $M$ can be defined in term of $\Delta^\prime$, which is written in the normalized form as $\Delta = \dfrac{\Delta^\prime}{\Delta_a}$. Here, $\Delta_a$ is the width of the attractor along the clipping direction~\cite{schrodertransient}. For numerical simulation, $\Delta \in [0,1]$ assigns the closed interval $[\text{global minima of the attractor}, \text{ global minima of the}\\ \text{attractor}+(\text{ global maxima of the attractor}-\text{ global minima of the attractor}) \times \Delta]$ as the mean field active region. Clearly, when $\Delta$ tends to $0+$, then neither of those $N$ oscillators get the suitable opportunity to interact among each other and as a consequence, only self-negative feedback is activated~\cite{dixit2019dynamicsfirst}. Besides, the global all-to-all interaction is established through mean field diffusive coupling for $\Delta \to 1-$. The value of $\Delta=1$ enables the entire phase-space as interaction active space.

\par To demonstrate our findings, the $d \times d$ diagonal matrix $\beta=\text{diag}(1,0,...,0)$ is considered by choosing $\beta_1=1$ and $\beta_j=0$ for $j=2,3,...,N$. Since, time-independent and time-varying diffusive interactions do not lead to stabilize the unstable stationary point of the system \cite{chowdhury2019extreme,9170822,mirollo1990amplitude,frasca2008synchronization,fujiwara2011synchronization} in general, our employed dynamic coupling is found to be beneficial to stabilize the networked oscillators from oscillatory behavior to death states. However, the nature of the emergent steady states may differ. The coupled oscillators may collapse into the existing stationary point of the uncoupled system. In the literature, this stationary point is known as amplitude death (AD) state \cite{ad_report}. On the other hand, it is also possible that coupled oscillators may converge into a new coupling-dependent steady state(s). This phenomenon is referred as oscillation death (OD) state \cite{od_report,chowdhury2020effect}. In the following section, we will explore the dynamics of the oscillators under the proposed dynamic coupling formalism. More precisely, we investigate the interplay between the control parameter $\Delta$ and the coupling strength $\varepsilon$, for which the discontinuous and irreversible transition from the oscillatory state to the death state occurs
in the ensemble of nonlinear oscillators. The equation \eqref{eq1} is integrated using the Runge-Kutta fourth-order (RK4) method for a time of $10^5$ units with a fixed integration time step $dt=0.01$ after removing enough transients of $10^6$ units throughout this letter.

\section{Results} \label{Results}
In this section, we discuss the explosive death state using the above coupling configuration. To do this, we consider three paradigmatic systems, namely Van der Pol oscillator (limit cycle), FitzHugh-Nagumo system (excitable system) and Lorenz system (chaotic system).
\subsection{Van der Pol Oscillator}\label{VDPresults}

\begin{figure}[h] 
	\centering
	\includegraphics[width=\linewidth]{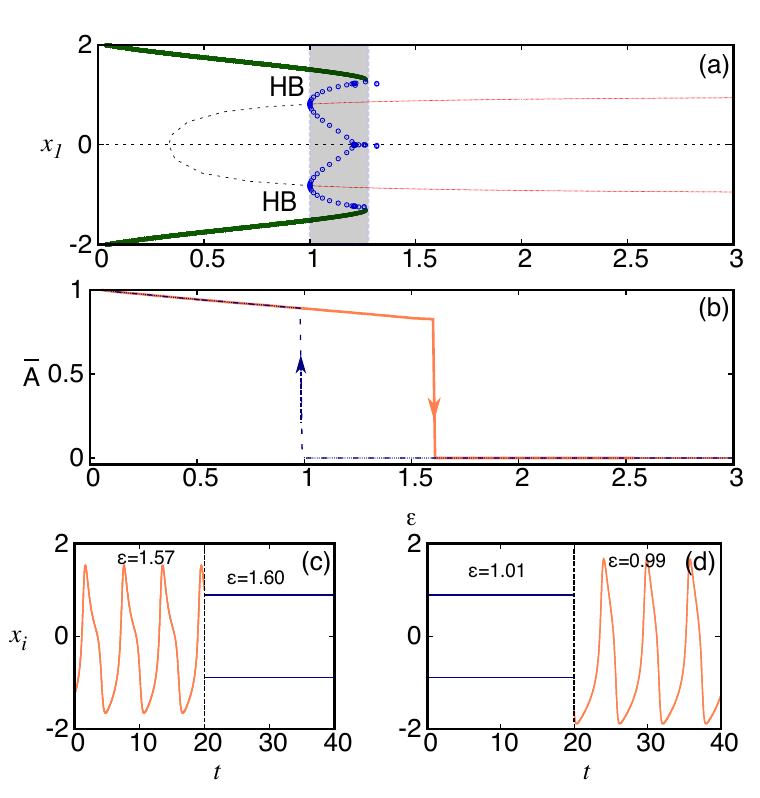}
	\caption{ (a) Bifurcation diagram of the VDP oscillator is plotted with respect to $\varepsilon$ at $\Delta=0$ using the software XPPAUT~\cite{ermentrout2002simulating}. Here, the red solid and black dotted lines show the stable and unstable steady states, respectively. Besides, green and blue circles represent stable and unstable periodic solutions, respectively. Here, HB denotes subcritical Hopf bifurcation point. (b) Forward and backward continuation of the order parameter $\bar{A}$ is shown here for $\Delta=0.5$ under the variation of the coupling strength $\varepsilon$ for $N=100$ coupled VDP oscillators. Time–series of the system \eqref{VDPmaineq1} near the transition point for (c) forward and (d) backward continuations is contemplated here at $\Delta = 0.5$. The strength of the damping parameter $b=3.0$ is set as fixed.	}
	\label{fig4VDP}
\end{figure}

We study a network of $N$ Van der Pol (VDP) oscillators \cite{kanamaru2007van} with dynamic mean field interaction as follows:
\begin{equation}
\begin{split}
\dot{x}_{j} &= y_{j} + \varepsilon({H(\overline{x})} \bar{x}-x_{j}), \\ 
\dot{y}_{j} &= b(1-x_{j}^2)y_{j}-x_{j},~~ j=1,2,\cdots, N.
\end{split}
\label{VDPmaineq1}
\end{equation}
Here, $\overline{x}=\frac{1}{N} \sum_{j=1}^{N} x_j$ and $(x_j, y_j)$ is the state variable of the $j$-th VDP oscillator. For positive values of the damping coefficient $b$, the VDP oscillator possesses a limit cycle. $\varepsilon$ stands for the interaction strength among interacted oscillators. All initial conditions are chosen randomly within $[-2,2] \times [-2,2]$.

\par For $\Delta=0.0$, the mean field interaction is completely inactive and thus, only negative self-feedback plays its role among those fully disconnected oscillators. To understand the qualitative changes of the system \eqref{VDPmaineq1} under self-feedback at $\Delta=0.0$ with respect to the $\varepsilon$, we plot the bifurcation diagram in Fig. \ref{fig4VDP}(a). This bifurcation diagram depicts that a symmetry breaking pitchfork bifurcation gives birth to {an $\varepsilon$-dependent nontrivial OD state $(x^*,y^*,-x^*,-y^*)$ at $\varepsilon \approx 0.33$. Here, $x^*=\pm \sqrt{1-\dfrac{1}{b\varepsilon}}$ and $y^*=\varepsilon x^*$. This OD state gains stability through subcritical Hopf bifurcation at $\varepsilon=1.0$, which is denoted by HB point in Fig. \ref{fig4VDP}(a). This HB point also gives birth to an unstable limit cycle. In the bifurcation diagram, the red solid and black dotted lines represent the stable and unstable steady states, while the green and blue circles signify stable and unstable periodic orbits, respectively. A shaded region is highlighted in this figure, where OD state coexists with a stable periodic orbit and an unstable limit cycle for $\varepsilon \in [1,1.265]$. The stable periodic orbit collides with the unstable periodic orbit at $\varepsilon=1.265$ and losses its stability. Thus, OD is the only remaining stable state beyond $\varepsilon=1.265$ and the bistable regime disappears as the stable limit cycle (green circle) becomes unstable.

\par To distinguish between the oscillatory state and steady state, the difference between the global maximum and minimum values of the attractor at a particular value of the coupling strength $\varepsilon$ is calculated and this is given by \cite{sharma2012amplitude}

\begin{equation}
a(\varepsilon) = N^{-1} \sum_{j=1}^N [ {\langle x_{j}^{max} \rangle}_{t} - {\langle x_{j}^{min} \rangle}_{t} ],
\label{op}
\end{equation}
after averaging it over the $N$ oscillators. Here, $\langle \cdots \rangle_t$ indicates the sufficiently long time average. The normalized average amplitude $\overline{A}$ is now treated here as an order parameter \cite{verma2017explosivescirep} and it is given by 
\begin{eqnarray}
\overline{A}=\frac{a(\varepsilon)}{a(0)}.
\label{contd}
\end{eqnarray}
\par Thus, this average amplitude parameter $\overline{A}$ lies within the closed interval $[0,1]$. The non-zero positive value of $\overline{A}$ reflects the oscillatory state of the system \eqref{eq1}. On the other hand, the death state is indicated through $\overline{A}=0$. In order to study the effect of dynamic interaction, $\Delta=0.5$ is considered where the interaction is active in the $50\%$ of the phase-space. In Fig.\ \ref{fig4VDP}(b), the variation of order parameter $\overline{A}$ at $\Delta=0.5$ with respect to variation
of the coupling strength $\varepsilon$ is plotted for both forward and backward continuations. The values of $\overline{A}$ are calculated following the steps provided in the Ref.\ \cite{verma2017explosivescirep}. The initial value of $\overline{A}$ at $\varepsilon=0$ is calculated using any random initial condition $(x_j(0), y_j(0))$ within $[-2,2] \times [-2,2]$. Then, $\varepsilon$ is gradually increased (in the case of forward continuation) adiabatically upto $\varepsilon=3.0$, i.e., the simulations are carried out for the next increased value of $\varepsilon$ using the final state of the state variable as the initial condition. The same method is also applied in the reverse direction (i.e., from $\varepsilon=3.0$ to $0.0$) for the backward continuation. 
 Fig.~\ref{fig4VDP}(b) reveals a sudden and discontinuous fall of the order parameter $\overline{A}=0$ in forward continuation at $\varepsilon=1.6$. 
 Similarly, the backward continuation also shows a sharp transition from $\overline{A}=0$ to a finite value at $\varepsilon=1.0$. These two transition points occur at different values of $\varepsilon$, and thus a hysteresis area is observed, which is the typical evocative for a first-order phase transition. The corresponding time-series of the system \eqref{VDPmaineq1} near both the forward and backward transitions is portrayed in Figs.~\ref{fig4VDP}(c) and \ref{fig4VDP}(d), respectively for $N=100$ oscillators. The oscillatory behavior is lost after the forward transition point $\varepsilon=1.6$ (before the backward transition point $\varepsilon=1$) and the system stabilizes to the stable OD state. It indicates the explosive transition of the amplitude as the strength of the coupling is changed. This fact is also supported in Fig.~\ref{fig4VDP}(b), where the values of $\overline{A}$ exhibit hysteresis revealing the coexistence of OD and stable periodic attractor for a given coupling strength within the interval $[1,1.6]$ of $\varepsilon$.

\begin{figure}[h]
	\centering
	\includegraphics[width=6 cm,height=3cm]{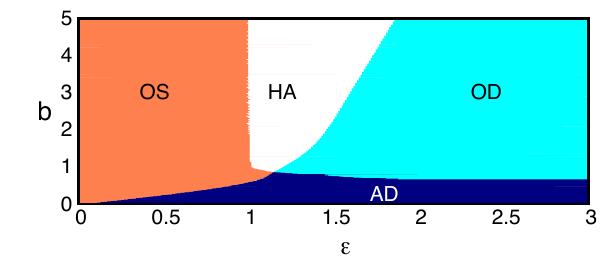}
	\caption{ Different dynamical domains of $N$ coupled VDP oscillators in the parameter plane $(\varepsilon-b)$. The regimes marked OS, HA, OD, and AD represent the oscillatory state, hysteresis area, oscillation death state, and amplitude death state, respectively. The other parameters are $\Delta=0.5$, and $N = 100$. $\Delta=0.5$ indicates that mean-field diffusive coupling is activated only when $\overline{x} \in [-2,0]$. Self feedback is the only active coupling in the complementary subspace.}
	\label{fig1VDP}
\end{figure}

\par The effect of the damping coefficient $b$ on the explosive transition of the $N=100$ coupled VDP system \eqref{VDPmaineq1} is depicted in Fig.~\ref{fig1VDP}. The phase diagram $(\varepsilon-b)$ is drawn at $\Delta=0.5$ by changing the values of $\varepsilon$ adiabatically in both forward and backward directions. In this figure $OS$,  $AD$,  and $OD$ describe oscillatory, amplitude death, and oscillation death states respectively. One can observe clearly that the system stabilized at $AD$ state $(x_j=0, y_j=0$, $j=1,2,...,N)$ with increasing coupling strength via second-order transition for $b<1.0$. A different scenario is observed for $b>1.0$, where the coupled system is stabilized to OD states via the first-order like transition. During this explosive death transition, a hysteresis region is found, where $OS$ and $OD$ solutions co-exist. The co-existence of $OS$ and $OD$ states is denoted as $HA$ in the Fig.~\ref{fig1VDP}. Interestingly, we can see that increment of parameter $b$ helps to enhance the hysteresis area in parameter space.

\par The backward transition point for this explosive transition can be computed using linear stability analysis around the $OD$ states of the system \eqref{VDPmaineq1} at $H(\overline{x})=0$. The OD states for this system \eqref{VDPmaineq1} at $H(\overline{x})=0$ are $x_{i}= x^{*}$, $y_{i}=y^{*}$, $\forall$ $i=1,2,\cdots,N$, where $x^{*} = \pm \sqrt{1-\frac{1}{b\varepsilon}}$ and $y^{*} = \varepsilon x^*$. For this stationary state, Jacobian matrix can be written as the block diagonal matrix $J \oplus J \oplus J \cdots \oplus J$ ($N$ times), where $J$ is the Jacobian matrix of the isolated system with only negative self-feedback at $(x^{*}, y^{*})$ given by

$$
J=
\begin{bmatrix}
-\varepsilon & 1 \\
1-2b\varepsilon & \dfrac{1}{\varepsilon}
\end{bmatrix}
$$

and the corresponding eigenvalues are

\begin{equation}
\lambda_{1,2}=\dfrac{1-\varepsilon^2 \pm \sqrt{\varepsilon^4-8b\varepsilon^3+6\varepsilon^2+1}}{2\varepsilon}.
\end{equation}

\par It gives the Hopf bifurcation point through which the OD solutions are stabilized at $\varepsilon_{HB} = 1$. A close inspection of the OD states and the corresponding eigenvalue analysis also help to detect the bifurcation point $\varepsilon_{PB}=\dfrac{1}{b}$, where the pitchfork bifurcation occurs. These Hopf bifurcation point $\varepsilon_{HB} = 1$ and the pitchfork bifurcation point $\varepsilon_{PB}=\dfrac{1}{b}$ match perfectly with our numerically found bifurcation diagram given in Fig.\ \ref{fig4VDP}(a). Note that the Hopf bifurcation point $\varepsilon_{HB} = 1$ is not only  independent of the number of oscillators $N$, but also it does not depend explicitly on $\Delta$. To validate our analytical finding and to understand the role of $\Delta$, we numerically investigate the phase diagram in the parameter plane $(\varepsilon-\Delta)$ for $b=3.0$. The interplay between $\Delta$ and $\varepsilon$ is portrayed in Fig.~\ref{fig3VDP}. The backward transition point fits exactly with our analytically calculated $\varepsilon_{HB} = 1$. $\Delta \sim 0.71$ creates two distinguished zones in this parameter space. For $\Delta \ge 0.71$, the system always settles down to stable oscillation for $b=3.0$ irrespective of the choice of the coupling strength $\varepsilon$. For $\Delta \ge 0.71$, the active interaction subspace $M$ always contain these OD states and thus, the mean field coupling always plays its role. Hence, the OD states of the system \eqref{VDPmaineq1} at $H(\overline{x})=0$ do not get the opportunity of being stabilized. For $\Delta<0.71$, an interval of $\varepsilon$ is found, where OD and oscillatory state coexist. This bistable region is highlighted as HA in Fig.~\ref{fig3VDP}. In a previous study of mean field interaction among identical VDP oscillators \cite{verma2017explosivescirep}, an intensity of mean field $Q$ $\in [0,1]$ is found to play an decisive role in the explosive death. Here, we have considered dynamic mean field interaction with control parameter $\Delta$ instead of density parameter $Q$, where interaction is on in a certain pre-defined subset of the state-space, otherwise remains off. This type of occasional interaction is relevant in various circumstances including transmissions of biological signals between synapses and
the communications of ant colonies in the processing of migration, as well as the seasonal interactions between predator–prey in the ecosystem \cite{sun2018inducing}. Robotic communication \cite{buscarino2006dynamical} and wireless communication systems are prominent specimens of such applications, where continuous interaction among the units is not always feasible.

\begin{figure}[h]
	\centering
	\includegraphics[width=6 cm,height=3cm]{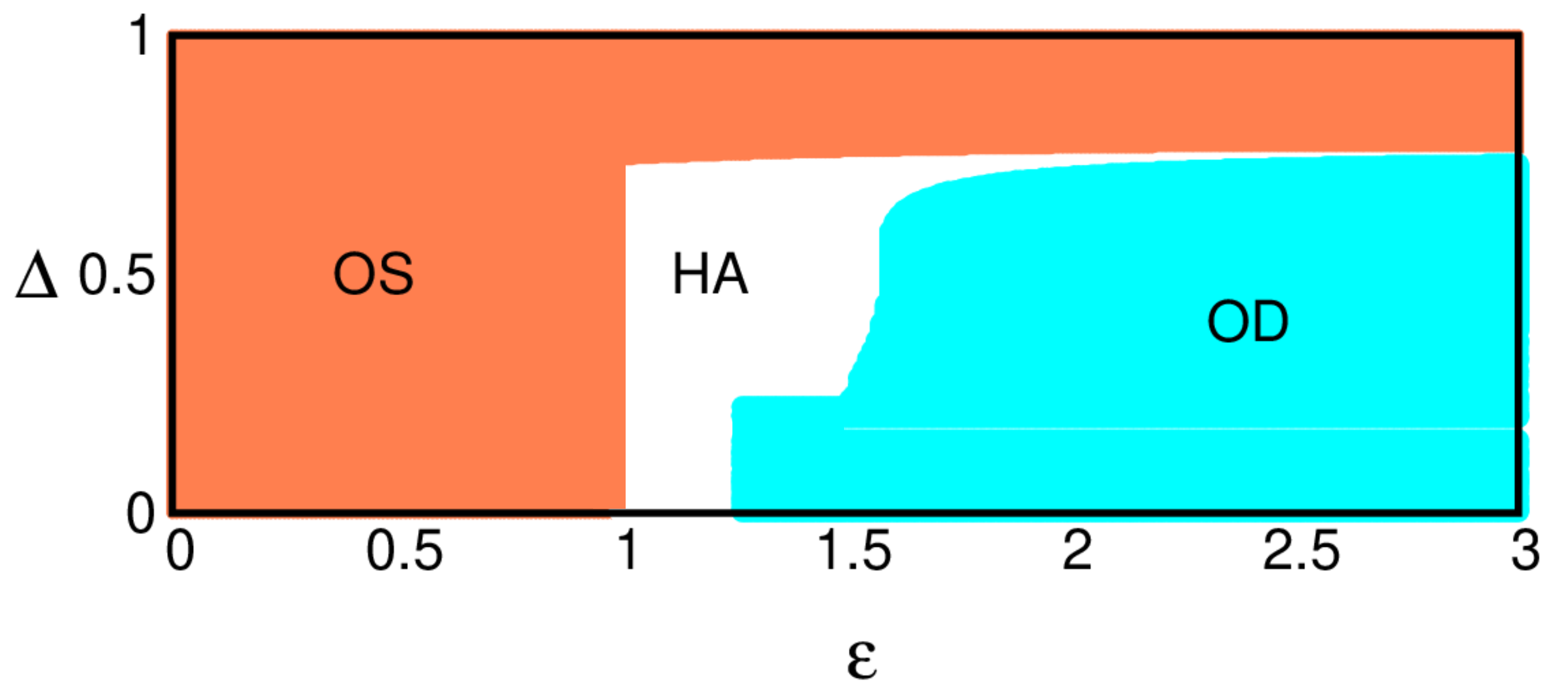}
	\caption{ Phase diagram of $N=100$ coupled VDP oscillators is represented here in the parameter plane $(\varepsilon-\Delta)$ for $b=3.0$. The bistable region is highlighted as HA separating the oscillatory state (OS) and the stationary state (OD) regime. The coupling strength $\varepsilon$ is varied adiabatically in both forward and backward directions to obtain this figure. The forward and backward transition points differ resulting in the explosive death phenomena.}
	\label{fig3VDP}
\end{figure}

\subsection{FitzHugh-Nagumo excitable system}\label{FHN}

\par To concur the universality of the explosive death in the coupled system \eqref{eq1} under proposed dynamic interactions, we adopt a more realistic neuronal model, namely, FitzHugh-Nagumo (FHN) excitable system \cite{fitzhugh1961impulses} for our study. The dynamical equation for the network consisting of FHN neurons under considered dynamic framework can be written as, 
\begin{eqnarray} \label{FHN}
\dot{x}_{j} &=& x_j(a-x_j)(x_j-1)-y_j + \varepsilon({H(\overline{x})} \bar{x}-x_{j}),\nonumber \\
\dot{y}_j &=& bx_j-cy_j.
\end{eqnarray}
Here, $x_j$ represents the trans-membrane voltage and the variable $y_j$ should model the time dependence of several physical quantities related to electrical conductances of the relevant ion currents across the membrane. In the FHN model, $x_j$ behaves as an excitable variable and $y_j$ acts as the slow refractory variable. We fix the parameters $a=-0.025, b=0.00652$ and $c=0.02$. The initial conditions are chosen from $[-0.4,1.4] \times [-0.4,1.4]$.

\par The coupled FHN system is investigated in the parameter plane $(\varepsilon-\Delta)$ as shown in Fig.\ \ref{fhnfig}(a). Interestingly, we report an explosive transition from oscillatory state to amplitude death state in certain parameter region. To the best of our knowledge, this novel explosive amplitude death phenomenon is reported for the first time in this letter. The earlier all perceived results on explosive death \cite{bi2014explosive,verma2017explosivescirep,verma2018first,verma2019explosive,verma2019explosivepre} is concerned only with the explosive transition from oscillatory states to OD states. Figure \ref{fhnfig} is drawn maintaining the same adiabatic process as discussed earlier. Just like our earlier observation with VDP oscillators, here we also find the hysteresis phenomenon, which is indicated in Fig.\ \ref{fhnfig}(a) as HA. The only difference is in the case of VDP oscillators, the HA regime contains oscillatory states and OD state, while in the case of FHN model, the AD state coexists with oscillatory states. The backward transition point is again found to be independent of $\Delta$. To calculate this backward transition point, the eigen values of the $N \times N$ Jacobian matrix $diag(J,J,...,J)$ around the AD state $(x_j=0, y_j=0,$~ $ j=1,2,...,N)$ of the system \eqref{FHN} is calculated at $H=0$. Here, $J$ is the Jacobian of the isolated FHN model at $(0,0)$. Thus,

$$
J=
\begin{bmatrix}
-a-\varepsilon & -1 \\
b & -c
\end{bmatrix}.
$$

\begin{figure}[h]
	\includegraphics[width=8 cm,height=5cm]{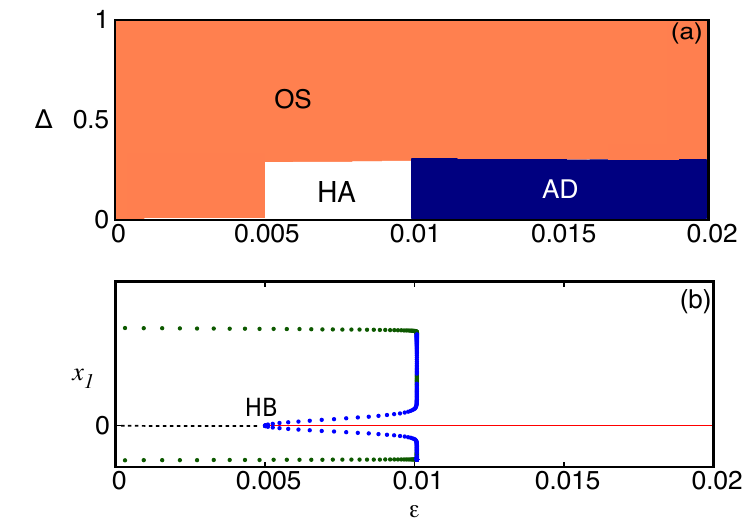}
	\caption{ (a) Phase diagram of $N=100$ coupled FHN system \eqref{FHN} in the $(\varepsilon-\Delta)$ plane. The other parameters are $a=-0.025, b=0.00652$ and $c=0.02$. Regions OS, HA, and AD indicate the oscillatory state, hysteresis area, and amplitude death state, respectively. In both forward and backward transitions, the system is integrated in an adiabatic fashion as discussed throughout the letter. The coexistence of AD and oscillatory states is responsible for the explosive transition. (b) Bifurcation diagram of the FHN oscillator at $H=0$. AD becomes stable at $\varepsilon=0.005$ and this is shown through red line. However, a stable periodic orbit (green circle) and an unstable periodic orbit (blue circle) coexist with AD for $\varepsilon \in [0.005, 0.01)$. The HB point indicates $\varepsilon=0.005$, before which AD is unstable (red line). }
	\label{fhnfig}
\end{figure}

\par The Hopf bifurcation occurs at $\varepsilon_{HB}=-(a+c)$, where the real part of the eigen values of $J$ become negative. For our choice of parameter values, the backward transition point is given by $\varepsilon_{HB}=0.005$. This transition point fits perfectly with our numercally derived results given in Fig.~\ref{fhnfig}. Clearly, this transition point is independent of the number of oscillators and the parameter $\Delta$. Figure \ref{fhnfig}(a) yields that the presence of a well pronounced hysteresis for $0 \le \Delta \leq 0.3$. For $\Delta > 0.3$, the system \eqref{FHN} exhibits the oscillatory behavior (OS) only. These trends are similar to those observed for coupled VDP oscillators suggesting generality of the explosive death phenomena.

\par In Fig.~\ref{fhnfig}(b), the bifurcation of the FHN oscillators is shown with respect to $\varepsilon$. Here, Hopf bifurcation point is denoted by HB, which matches perfectly with our analytically calculated value. The red and black lines represent stable and unstable steady state, while the green and blue circles represents stable and unstable limit cycle, respectively. The unstable origin stabilizes at $\varepsilon=0.005$. Unstable periodic orbit originates too at this backward transition point through subcritical Hopf bifurcation. This unstable periodic orbit, stable origin and a stable periodic orbit all co-exist for $\varepsilon \in [0.005,0.01)$. At $\varepsilon=0.01$, the stable periodic orbit collides with the unstable periodic orbit and loses their stability and eventually disappears. The origin is the only stable state for $\varepsilon>0.01$ and $0 \le \Delta \leq 0.3$. Coexistence of AD and oscillatory state gives rise to first order like transition with hysteresis.

\subsection{Lorenz system}\label{Lorenzresults}

\par To speculate this irreversible mechanism, we now consider the case of $N$ coupled chaotic Lorenz oscillators \cite{lorenz1963deterministic} interacting via dynamic mean field interaction. The mathematical equations representing the network dynamics are, 
\begin{equation}
\begin{split}
\dot{x}_{j} &= \sigma(y_{j}-x_{j}) + \varepsilon({H(\overline{x})} \bar{x}-x_{j}), \\ 
\dot{y}_{j} &= (\rho-z_{j})x_{j}-y_{j}, \\
\dot{z}_{j} &= x_{j}y_{j}-\beta z_{j},   
\end{split}
\label{lorenzeq1}
\end{equation}
where $j=1,2, ... ,N$ is the index of oscillators. The system parameters are chosen as $\sigma=10$, $\rho=28$, and $\beta=\dfrac{8}{3}$ for which an individual system is in chaotic state. Random initial conditions are chosen within $[-30, 30] \times [-30, 30] \times [-30, 30]$.

%

\begin{figure}[h]
	\centering
	\includegraphics[width=6 cm,height=3cm]{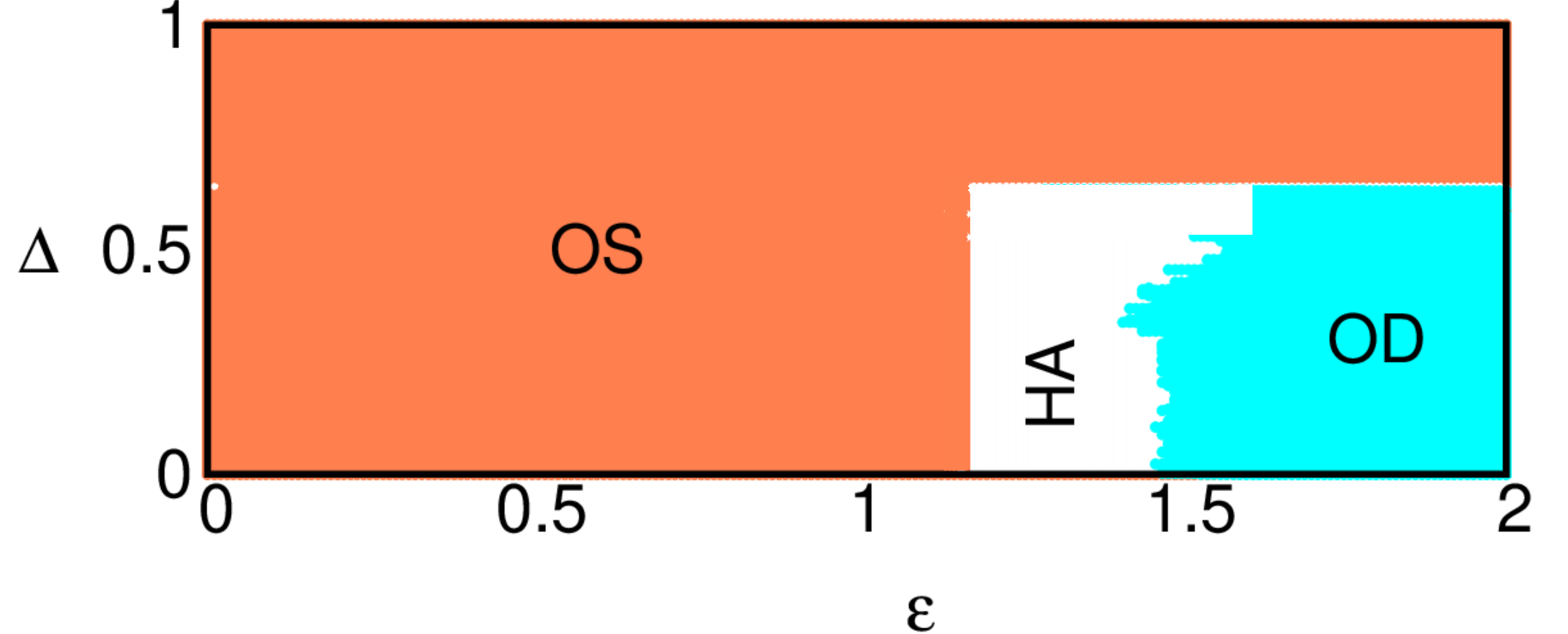}
	\caption{ Interplay of $\varepsilon$ and $\Delta$ generates different dynamical domains including the oscillatory state (OS), hysteresis area (HA) and oscillation death states (OD). The simulation is carried out for $N=100$ coupled Lorenz oscillators with adiabatic increament of $\varepsilon$ both in forward and backward directions. Clearly, the backward transition point ($\varepsilon=1.171$) is independent of $\Delta$. }
	\label{lorenzfig}
\end{figure}
%

\par By adiabatically changing $\varepsilon$ in both directions with step size $\delta \varepsilon = 0.01$, we are able to capture the different emerging behaviors of coupled Lorenz oscillators in the $(\varepsilon-\Delta)$ parameter space (Fig.~\ref{lorenzfig}). The Jacobian matrix at $H=0$ corresponding to the OD states $x_{i}= x^{*}$, $y_{i}=y^{*}$, $z_{i}=z^{*}$ ($i=1,2,...,N$), where $x^{*} = \pm \sqrt{\dfrac{\beta \sigma (\rho-1)-\beta \varepsilon}{\sigma+\varepsilon}}$, $y^{*} = \dfrac{(\sigma+\varepsilon)x^{*}}{\sigma}$ and $z^{*}=\dfrac{x^{*} y^{*}}{\beta}$ can be represented by a $N \times N$ block circulant matrix circ$(J,\bf{0},\cdots,\bf{0})$ with
$$
J=
\begin{bmatrix}
-\sigma-\varepsilon & \sigma & 0\\
\rho-z^{*} & -1 & -x^{*}\\
y^{*} & x^{*} & -\beta
\end{bmatrix}.
$$

\par Negativity of the real part of all eigen values of this block circulant matrix determines the backward critical point at $\varepsilon=1.171$, which matches with our numerical simulation given in Fig.~\ref{lorenzfig}. This transition point is independent of $N$, i.e., the number of oscillators present in the system. As we vary $\varepsilon$ in the both directions with step size $\delta \varepsilon = 0.01$ adiabatically, a first order like phase transition with hysteresis is found within the interval $\varepsilon \in [1.18, 1.65]$ and $\Delta \in [0, 0.64]$. This hysteresis regime with distinct forward and backward transitions is emphasized by marking this region as HA in Fig.~\ref{lorenzfig}. Outside of this HA region, either oscillatory behavior (OS) or OD states are stabilized as depicted with this figure of $(\varepsilon-\Delta)$ parameter space. However, one should notice that the backward transition point remains unchanged with variation of $\Delta$.

\section{Conclusion}\label{conclusion}

The present work has demonstrated that apart from the continuous interaction, dynamic interaction leads to an explosive death transition in coupled nonlinear oscillators. We have studied explosive first order like dynamical transition in nonlinear oscillators interacting through the state-space dependent dynamic coupling. Using limit-cycle and chaotic oscillators interacting through dynamic interaction, we have shown that the system possesses two distinct states, viz.\ the oscillatory state and the steady state and the transition between these two states is irreversible, abrupt, and associated with the presence of a hysteresis region where such states co-exist. The hysteresis region crucially depends on the interaction region in state-space as well as the coupling strength. The analysis performed here may help to understand the effect of dynamic interaction leading to the interesting collective dynamical behavior of coupled systems and its relevance with the occurrence of such states in many natural systems.

\medskip
\acknowledgments
	
MDS acknowledges financial support (Grant No.\ EMR/2016/005561 and INT/RUS/RSF/P-18) from Department of Science and Technology (DST), Government of India, New Delhi. S.N.C. would  like  to  acknowledge  the  CSIR  (Project No. 09/093(0194)/2020-EMR-I) for financial assistance.

\bibliographystyle{iopart-num}
\bibliography{ref}

\end{document}